\documentclass[12pt]{article}
\usepackage{graphicx}
\usepackage{verbatim}
\usepackage{amsmath,amsthm,amsfonts}
\usepackage{amssymb}
\usepackage{hyperref}
\usepackage{caption}
\DeclareCaptionFont{10pt}{\fontsize{11pt}{12pt}\selectfont}
\theoremstyle{definition}

\newtheorem{def-theorem}[theorem]{Definition-Theorem}

\newcommand{\be}{\begin{equation}}
\newcommand{\ee}{\end{equation}}
\newcommand{\bea}{\begin{eqnarray}}
\newcommand{\eea}{\end{eqnarray}}
\newcommand{\beas}{\begin{eqnarray*}}
\newcommand{\eeas}{\end{eqnarray*}}
\newcommand{\ba}{\begin{array}}
\newcommand{\ea}{\end{array}}

\begin{document}

\begin{titlepage}
\hfill
\vbox{
    \halign{#\hfil         \cr
           } 
      }  
\vspace*{15mm}
\begin{center}
{\Large \bf Building up spacetime with quantum entanglement II: \vskip 0.05 in It from BC-bit}

\vspace*{8mm}
\vspace*{1mm}
Mark Van Raamsdonk
\vspace*{0.5cm}
\let\thefootnote\relax\footnote{mav@phas.ubc.ca}

{\it Department of Physics and Astronomy,
University of British Columbia\\
6224 Agricultural Road,
Vancouver, BC, V6T 1Z1, Canada \\
}

\vspace*{0.5cm}
\end{center}
\begin{abstract}
In this note, we describe how collections of arbitrary numbers of ``BC-bits,'' distinct non-interacting quantum systems each consisting of a holographic boundary conformal field theory (BCFT), can be placed in multipartite entangled states in order to encode single connected bulk spacetimes that approximate geometries dual to holographic CFT states. The BC-bit version of a holographic CFT state corresponds to a geometry that can be made arbitrarily similar to the associated CFT-state geometry within a ``causal diamond'' region defined by points that are spacelike separated from the boundary time slice at which the state is defined. These holographic multi BC-bit states can be well-represented by tensor networks in which the individual tensors are associated with states of small numbers of BC-bits.
\end{abstract}

\end{titlepage}

\subsubsection*{Introduction}

There is by now a significant amount of evidence that the emergence of spacetime in holographic models of quantum gravity is directly linked to the entanglement between the underlying degrees of freedom (see, for example \cite{Maldacena:2001kr, Ryu:2006bv, Swingle:2009bg, VanRaamsdonk:2009ar,VanRaamsdonk:2010pw, Maldacena:2013xja}, or \cite{VanRaamsdonk:2016exw} for a review). Nevertheless, this basic phenomenon is somewhat obscured by the continuous nature of the conformal field theory (CFT) systems that encode the spacetimes. In these examples, part of the spacetime structure (the fixed asymptotic behavior) is directly related to the continuous geometrical space upon which the CFT is defined. Also, the local degrees of freedom interact strongly with those around them, and completely disentangling the various parts requires an infinite amount of energy. There are interesting toy models for holography involving tensor-network states of collections of individual qubits or other discrete elementary subsystems \cite{Swingle:2009bg, Pastawski:2015qua, Hayden:2016cfa}, but it still not clear how closely the network geometries associated to these states are related to the actual geometries in holographic models.

In this note, we introduce a new framework for holography in which the fundamental degrees of freedom are a large collection of elementary systems (boundary conformal field theories which can be thought of as ``bits'' of the original CFT) which do not interact with each other. States of a holographic CFT can be replaced by entangled states of this discrete system so that the system still describes a single connected spacetime that is close to the original one dual to the CFT state. Furthermore, in the new description, we show that there is a natural way to represent the state with arbitrary precision using a type of tensor network.

\subsubsection*{BC bits}

The motivation for our construction is the idea of cutting up a holographic conformal field theory into a large number of non-interacting pieces (e.g. using a CFT ``jigsaw'') but putting the pieces into a quantum state that approximates a state of the original CFT encoding some geometry. To make this precise, we need to define what me mean by a CFT living on a piece of space with a boundary. Roughly, we need to describe the boundary conditions for the fields at the edges. For any CFT, there are various choices of boundary conditions that are consistent. A special subset of these preserve conformal invariance\footnote{For example, the vacuum state of the CFT on a half space with these boundary conditions at the edge preserves an $SO(d-1,2)$ of the $SO(d,2)$ conformal symmetry.} and define what is known as a boundary conformal field theory or BCFT (see, for example, \cite{Cardy:2004hm}).

A set of ``Boundary-CFT bits'' or ``BC-bits'' associated with a CFT on some spatial geometry $M$ is a collection of BCFTs defined on a set of ``sanded'' pieces $\tilde{M}_i$ of $M$. Here, $\{\tilde{M}_i\}$ are defined by cutting $M$ into a set of simply-connected pieces $\{M_i\}$ and ``sanding the edges,'' i.e. $\tilde{M}_i$ is a large subset of the interior of $M_i$ with a smooth boundary. This is shown in figure \ref{fig:SphereBits}a) - c). Each BCFT is defined from the original CFT with the same choice boundary conditions, so the BC-bit system is specified by the choice of $\{\tilde{M}_i\}$ and the choice of boundary condition.

Our goal below will be to consider some state of the CFT that corresponds to a smooth geometry and associate to this a state of the BC-bits that captures the same qualitative features. We will argue that the new state can also be associated with a smooth geometry that is closely related to the original one.

\begin{figure}
\centering
\includegraphics[width=140mm]{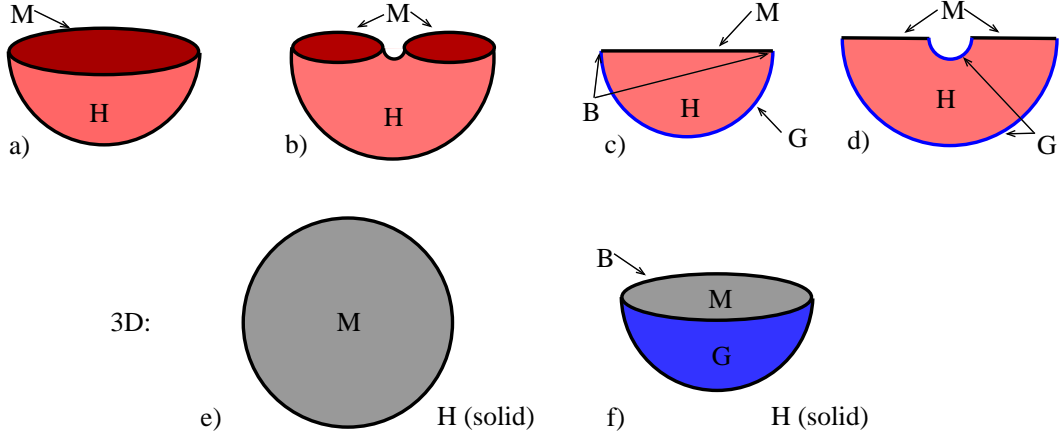}
\caption{Euclidean path integrals defining a) a state of a 2D CFT on a circle b) an entangled state of two 2D CFTs each on a spatial circle c) a state of a 2D BCFT on an interval d) an entangled state of two 2D BCFTs e) a state of a 3D CFT on a sphere f) a state of a 3D BCFT on a disk.}
\label{fig:PIsimple}
\end{figure}

\subsubsection*{Entangling BC-bits via Euclidean path integral}

\begin{figure}
\centering
\includegraphics[width=120mm]{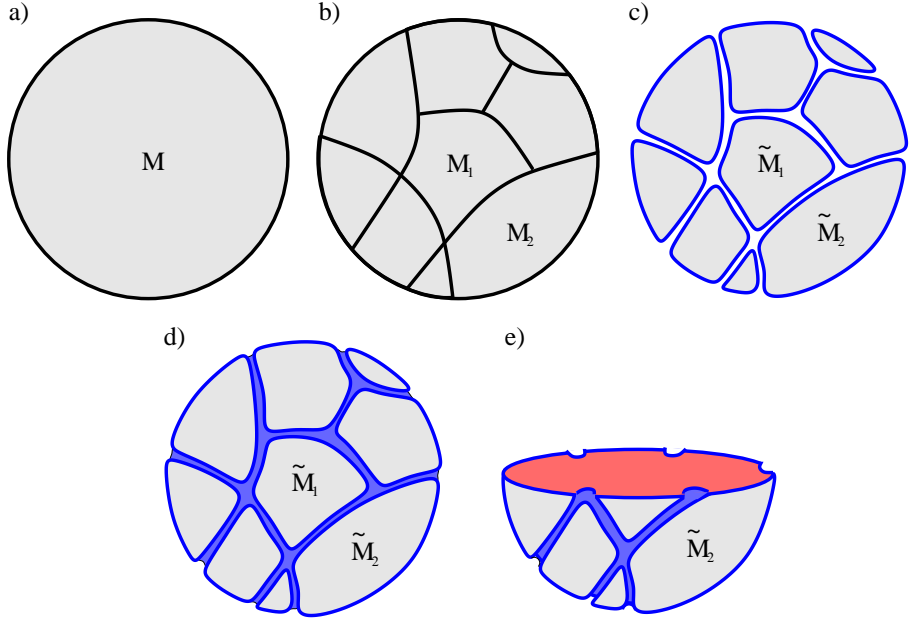}
\caption{a)-c) BC-bits defined from a 2+1 dimensional CFT on a spatial sphere. d) Geometry $\tilde{H}$ used to define a state of the BC-bits. e) Cross section of $\tilde{H}$.}
\label{fig:SphereBits}
\end{figure}
The Euclidean path integral gives a mechanism to define a quantum state of a CFT on a spatial manifold $M$ given a spatial manifold $H$ of one higher dimension with boundary $M$. The probability amplitude in this state for a having a particular configuration of fields on $M$ is given by the CFT path integral\footnote{This is an average over the possible field configurations in the Euclidean version of the CFT, weighted by the exponential of minus the Euclidean action.} on $H$ subject to the condition that the fields take the specified values on $M$. For holographic CFTs, this construction gives a natural way to define states that encode dual spacetimes with a good classical description (see, for example, \cite{Skenderis:2008dh, Botta-Cantcheff:2015sav,Marolf:2017kvq} and references therein). We can change the spacetime we are describing by varying the interior geometry of $H$ and adding sources for various CFT operators on the interior of $H$; for example, arbitrary linearized perturbations about pure AdS can be obtained by choosing the right geometry and sources \cite{Marolf:2017kvq}.

We can similarly define the state of a BCFT on a spatial manifold $M$ with boundary $B$ given a higher-dimensional manifold $H$ with boundary $M \cup G$, where $G$ is also bounded by $B$, as shown in figures \ref{fig:PIsimple}c and \ref{fig:PIsimple}f. In this case, we take the boundary conditions at $G$ to be those of the BCFT we are considering.

We can use the Euclidean path integral to define natural entangled states of non-interacting CFTs or BCFTs by taking the $H$ to be a connected geometry with a disconnected boundary $M$ (see, for example \cite{Balasubramanian:2014hda}). For example, the thermofield double state for a pair of CFTs defined on spatial spheres is defined by the path integral on a cylinder defined by the spatial sphere times an interval (figure \ref{fig:PIsimple}b). We will use this idea to define an entangled state of our BC-bits by taking $H$ to be a connected geometry whose boundary is the collection $\{\tilde{M}_i\}$.

More specifically, we would like to define a state of the BC-bits that is related to some state of the original CFT on $M$ defined by the path integral over $H$. To do so, we define a geometry $\tilde{H}$ obtained from $H$ by removing smooth ``grooves'' at the surface\footnote{A mathematical ``router'' is the appropriate tool here.} so that the part of the boundary remaining is $\tilde{M}_i$. This is depicted in figure \ref{fig:SphereBits}d for a 2+1 dimensional CFT and in figure $3 \tilde{\rm b}$ for a 1+1 dimensional CFT. The boundary of $\tilde{H}$ is $\left\{ \cup_i \tilde{M}_i \right\} \cup G$, where $G$ corresponds to the surface of the grooves. We can now define a state of the BC-bits by performing the Euclidean path integral for our BCFT over the geometry $\tilde{H}$, with the appropriate boundary conditions imposed at $G$.

The state $|\Psi_{\tilde{H}} \rangle$ and the possible dual geometry will depend on the details of $\tilde{H}$. However, we will now argue that if we have been sufficiently gentle with our carpentry tools, i.e. if $\{\tilde{M}_i\}$ is close enough to $\{M_i\}$ and $\tilde{H}$ is close enough to $H$, that the state $|\Psi_{\tilde{H}} \rangle$ of our BC-bits encodes a geometry that is closely related to the original geometry described by the CFT state $|\Psi_{H} \rangle$.

\begin{figure}
\centering
\includegraphics[width=120mm]{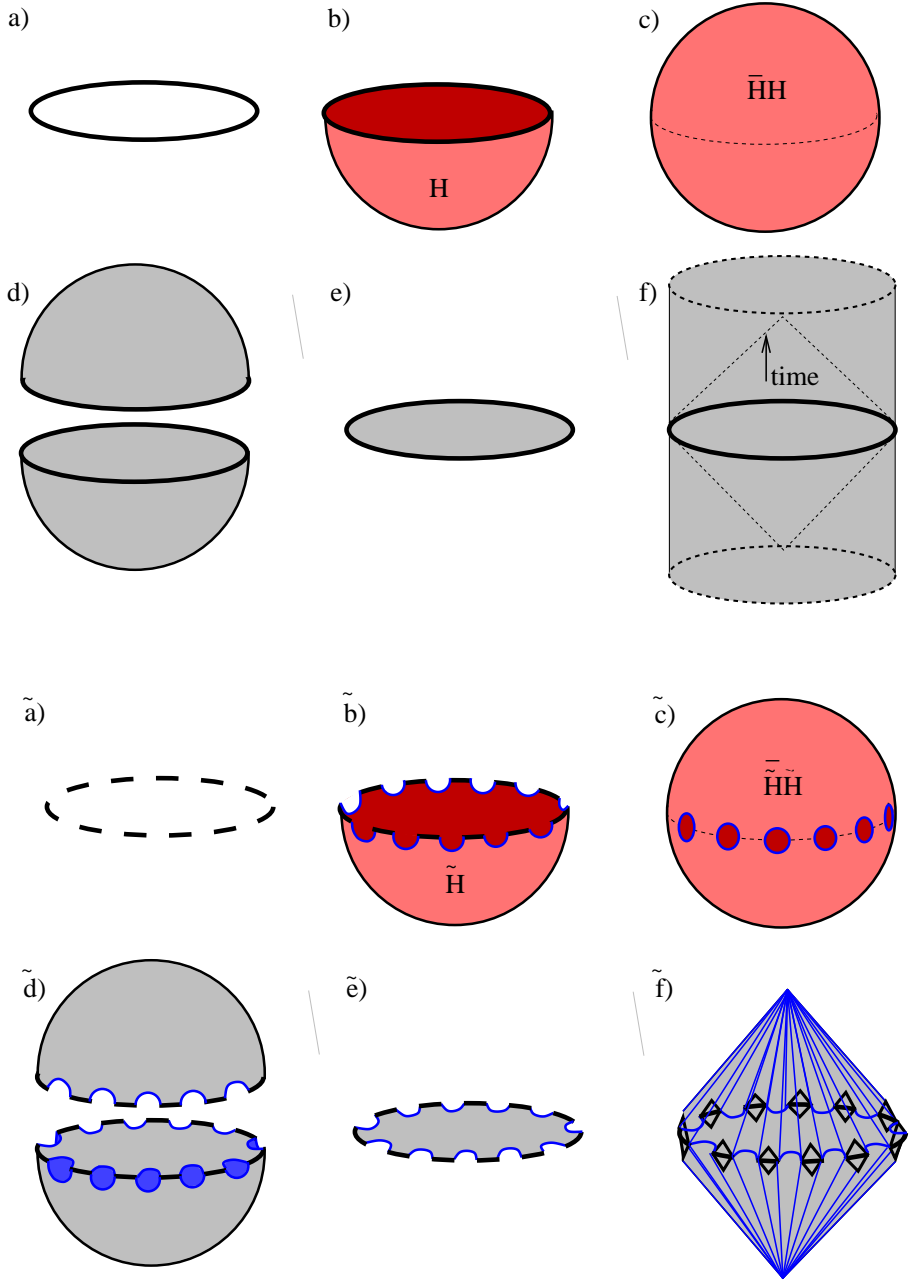}
\caption{Lorentzian geometries from path-integral states: a) A CFT on a circle. b) Euclidean path integral defining a holographic CFT state. c) Path integral used to compute observables in this state. Operators can be inserted on dashed lines. d) Euclidean gravity solution corresponding to this path integral. e) Spatial slice at time-symmetric point serves as initial data for Lorentzian solution. f) Lorentzian solution associated with our state. The interior of the causal diamond (dashed lines) is the part encoded by the CFT state at $t=0$. $\tilde{a}$) - $\tilde{f}$) Equivalent construction for the BC-bit states. Each BC-bit is a boundary CFT on an interval.}
\label{fig:PathInt}
\end{figure}

\subsubsection*{Spacetimes dual to Euclidean path integral states}

Let us now recall how to understand the spacetime geometries encoded by holographic CFT states described by a Euclidean path integral. According to the AdS/CFT correspondence, features of the encoded geometry can be deduced from the CFT by evaluating the expectation value of various local and nonlocal observables in the CFT state. For our states, these expectation values are computed using a path integral over a surface $\bar{H}H$ that is obtained by gluing $H$ to its mirror image $\bar{H}$ along $M$, as seen in figure \ref{fig:PathInt}c and \ref{fig:PathInt}$\tilde{c}$.\footnote{Any complex sources on $H$ should be conjugated in $\bar{H}$, but we will restrict to the case of real sources; the resulting geometry will then have a time-reversal symmetry. We can restrict to geometries $H$ so that the resulting space $\bar{H}H$ is smooth.} The operators of interest are inserted along the junction. By the AdS/CFT correspondence this CFT path integral corresponds to a gravitational path integral which is dominated by a single Euclidean geometry $X_{H}$, obtained by solving the gravitational equations with boundary conditions that $X_{H}$ is asymptotically AdS with boundary geometry $\bar{H}H$.\footnote{The asymptotic behaviour of other fields in the geometry is fixed by the sources for the corresponding operator in the path integral action.}

The Lorentzian spacetime geometry associated with our CFT state is simply related to the Euclidean geometry $X_{H}$. The geometry $X_{H}$ has a reflection symmetry inherited from the geometry of $\bar{H}H$. The surface left invariant under this symmetry has a geometry $(X_{H})_0$. To find the spacetime associated with our state, we use the geometry $(X_{H})_0$ (and the condition that time-derivatives of fields vanish here) as initial conditions for the real-time gravitational equations. The solution is a spacetime $X^L_{H}$ corresponding to our state. More formally, we can understand $X^L_{H}$ as an analytic continuation of $X_{H}$. The CFT state at $t=0$ strictly encodes only the region of this spacetime that is spacelike separated from the $t=0$ slice at the boundary,\footnote{This is the domain of dependence of the region $(X_{H})_0$.} since the spacetime outside this region can be altered by changes to the CFT Hamiltonian before or after $t=0$. For example, a modification to the CFT Hamiltonian at $t=\epsilon$ corresponds to a boundary source in the gravity picture whose effects propagate forward causally from the $t=\epsilon$ boundary slice. Thus, the geometry directly encoded by the CFT state is this causal diamond region, also known as the Wheeler-DeWitt patch for the $t=0$ boundary time slice.\footnote{This connection between the quantum state of a holographic theory at a particular time and the Wheeler-DeWitt patch of the dual geometry associated to that time has been emphasized at various times in the past, for example in the recent conjecture that the gravitational action integrated over the Wheeler-DeWitt patch corresponding to some boundary time provides a measure of complexity of the dual CFT state at that time \cite{Brown:2015bva}}.

The crux of our subsequent argument will be that despite the significant differences between the original CFT and the collection of BC-bits as physical systems, the geometry $\tilde{H}$ used to define the BC-bit state is a small perturbation to the geometry $H$ used to define the CFT state. Thus, we might expect that the procedure we have just outlined gives rise to a spacetime dual to the BC-bits that is almost the same as the spacetime dual to the CFT state.

\subsubsection*{Geometry of the BC-bit states}

We would like to understand how the Euclidean gravity solution corresponding to a BCFT on $\bar{\tilde{H}}\tilde{H}$ differs from that corresponding to the CFT on $\bar{H} H$ (figure 3b vs figure $3 \tilde{\rm b}$). The main obstacle here is understanding how the presence of a boundary in $\bar{\tilde{H}}\tilde{H}$ (geometrically described as the space $G$ glued to a mirror image of itself along $B$) affects the gravity calculation. This question was considered originally by Karch and Randall in \cite{Karch:2000gx} and later in more detail by Takayanagi in \cite{Takayanagi:2011zk}. As discussed in those papers, if the BCFT state has a geometrical dual $X_{\tilde{H}}$, this dual must itself have a boundary component in addition to the asymptotically AdS boundary with boundary geometry $\tilde{H}$. What could this boundary be physically in a UV complete gravity theory? One possibility is that we have some compact internal dimensions that degenerate somehow, in the way that the Kaluza-Klein circle shrinks to zero size to for the edge of a bubble-of-nothing geometry. The boundary could also or alternatively involve some explicit branes of the microscopic theory. Examples of this sort have been described explicitly for higher dimensional CFTs in \cite{Chiodaroli:2011fn, Chiodaroli:2012vc}. We expect that the specific way such a boundary is realized will depend on the boundary conditions we have chosen for our BCFT.

\begin{figure}
\centering
\includegraphics[width=120mm]{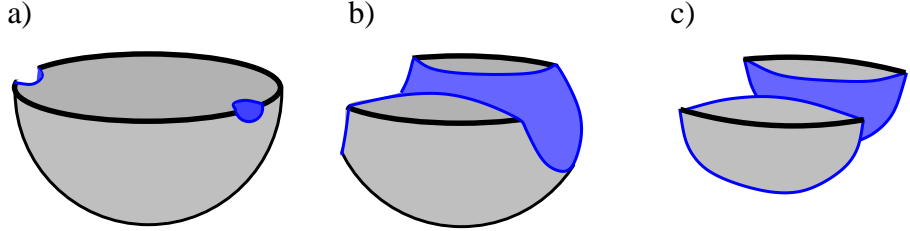}
\caption{Euclidean geometries associated with path integrals for states of two BC-bits, using the end-of-the-world brane model for holographic BCFTs. Each shows half of the Euclidean geometry, up to the slice that becomes the initial data for the Lorentzian geometry. a) When $\tilde{H}$ is sufficiently close to $H$, the ETW branes are disconnected and localized near the corresponding boundary components. The Lorentzian initial data slice is connected and similar to that for the single CFT path integral state defined by $H$. b) As the modifications defining $\tilde{H}$ become too severe, we can have a  transition in which the ETW brane topology changes. Here, the Lorentzian initial data slice becomes disconnected, though there will still be entanglement between the matter in the two components of spacetime. c) A path integral defining a pure state of the same two BC-bits as in b). This corresponds to two disconnected spacetimes with no entanglement between the matter in the different components.}
\label{fig:PhaseTrans}
\end{figure}

As a simple model, we can make use of the suggestion in \cite{Karch:2000gx,Takayanagi:2011zk} to introduce an explicit end-of-the-world (ETW) brane with constant tension and Neumann boundary conditions. In that case, as long as a given boundary region of $\tilde{H}$ is small compared to other geometrical features of $\tilde{H}$ (such as the distance to other boundary components) the ETW brane ending on that boundary component stays localized to the vicinity of that boundary component, as depicted in figure \ref{fig:PathInt}$\tilde{d}$ and \ref{fig:PhaseTrans}a (blue surfaces). In the limit where these boundary components become small, the ETW brane also becomes localized to the asymptotic region of the geometry near this boundary component. The ETW brane may source bulk fields and affect the rest of the geometry, but for a fixed location away from the ETW brane, these effects will become negligible in the limit where the boundary component is taken small. By contrast, if the boundary region of $\tilde{H}$ becomes too large, we can have phase transitions where the ETW brane topology changes, as shown in figure \ref{fig:PhaseTrans}. In this case, we might end up with a spacetime that has disconnected components.

For 1+1 dimensional CFTs, we can give a more direct argument that doesn't rely on a particular holographic model for BCFTs. In this case, $\bar{\tilde{H}}\tilde{H}$ is a two-dimensional Euclidean geometry, as in the example of figure \ref{fig:PathInt}$\tilde{c}$. Here, we can take each component of $G$ to be a small circle as displayed in that figure. If the circle is sufficiently small relative to the distance to other boundary components and operator insertions, we expect (as in the operator product expansion) that its insertion is equivalent to the insertion of a sum of operators $\sum_i c_i {\cal O}_i$, where $c_i$ depends on the size of the disk. To find the $c_i$, we can consider the circular boundary inserted into the path integral on a disk that defines a state of the CFT on the circular boundary of our disk. By a conformal transformation, this path integral gives an equivalent state the path integral on a finite cylinder, where one end is the circle where our CFT lives and the other end is the circle where we apply our boundary conditions. The limit where the circle in the original picture becomes small corresponds to a limit where the cylinder becomes long. But in the limit of an infinitely long cylinder, the state we define is the vacuum state. Thus, regardless which boundary conditions we are considering to define our BCFT, we can say that the coefficients $c_i$ for operators which are not the identity operator go to zero in the limit where the circle becomes small.

According to these arguments, the Euclidean geometry associated with $\bar{\tilde{H}}\tilde{H}$ should be almost the same as that associated with $\bar{H} H$. But this is not quite true for the corresponding Lorentzian geometries. The reason is that no matter how small the boundary components of $\bar{\tilde{H}}\tilde{H}$ are, they still change the asymptotic geometry of the slice that serves as the initial data for our Lorentzian evolution (see figure \ref{fig:PathInt}$\tilde{d}$). Thus, the state we are defining, considered as a state of the original CFT, will always have infinite energy. In the OPE language, the infinite energy is associated with the fact that we are acting with a local operator on the vacuum state; in this case, no matter how small the coefficient of this operator is in the superposition, the average energy for the full state is still infinite. The result is that in the Lorentzian picture, the initial data slice is almost the same as that for the original CFT state, but because of the differences in asymptotics at the boundary, we can have some type of shock wave evolving forward and backward from each boundary component we have introduced. In the ETW brane picture, we can understand this as a Lorentzian ETW brane whose worldvolume is part of a hyperboloid (the analytic continuation of the hemispherical ETW branes in the Euclidean picture of figure \ref{fig:PathInt}$\tilde{d}$. The Lorentzian spacetime is depicted in figure \ref{fig:PathInt}$\tilde{f}$.

In a limit where we have very many BC-bits and very many small boundary components, this shockwave or ETW brane will propagate outward from the full asymptotic boundary of the initial data slice. Thus, the BC-bit version of a holographic state faithfully reproduces the interior of the Wheeler-DeWitt patch associated with the boundary time at which the state is defined, but the geometry generally will not smoothly continue to the past and futire of this patch.

\subsubsection*{Tensor networks for holographic BC-bit states}

The BC-bit construction of holographic states bears a closer resemblance to tensor network toy models of holography \cite{Swingle:2009bg, Pastawski:2015qua, Hayden:2016cfa} in that we have explicit multipart entangled states of a discrete set of non-interacting constituents. We will now see that as in the toy models, our states may be represented arbitrarily well by a type of tensor network, where the tensors correspond to states of small numbers of BC-bits. Our construction involves additional small changes to $\tilde{H}$, this time carried out using a mathematical ``drill''.

We have argued previously that the introduction of a small boundary component to $\tilde{H}$ has a vanishingly small effect on the dual Euclidean geometry in the limit that the size goes to zero, except very close to the insertion locus. Thus, if we introduce additional boundary components away from $M$, these should have a negligible effect on the BC-bit state as the new boundary components are taken to be infinitesimal.

\begin{figure}
\centering
\includegraphics[width=120mm]{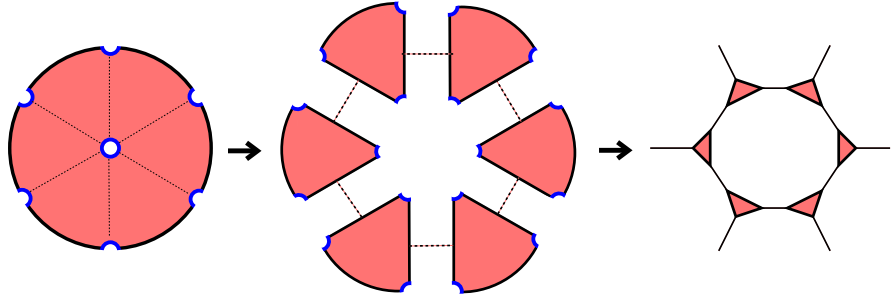}
\caption{a) An extra circular boundary component is added to the interior of $\tilde{H}$ in the path integral for a six BC-bit state. b) The path integral can now be decomposed into a product of path integrals defining three BC-bit states. The field configurations for the BC-bits connected by dashed lines are equated and integrated over - the path integral equivalent of the pair projection that connects a tensor network. c) The tensor network representation of the state.}
\label{fig:TN}
\end{figure}

\begin{figure}
\centering
\includegraphics[width=110mm]{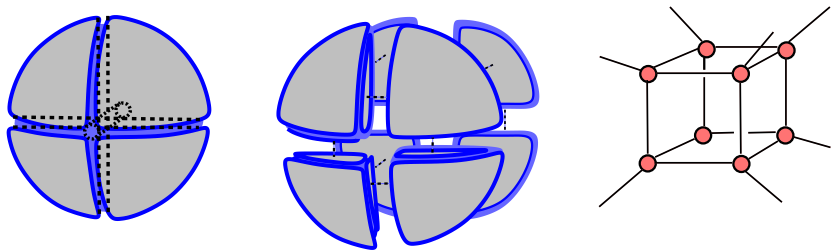}
\caption{Tensor network for path-integral state of eight BC-bits. In the first step, the state is modified by drilling narrow holes through the ball along the axes, enlarging the boundary of $\tilde{H}$. In general, for three dimensional $\tilde{H}$ the modifications necessary in the first step involve introducing a network of narrow tunnels through $\tilde{H}$.}
\label{fig:TN3D}
\end{figure}

By introducing these extra boundaries (we can imagine drilling through $\tilde{H}$ with a narrow drill bit), it is possible to represent the new state as a tensor network, by decomposing the path integral into chunks, as shown in figures \ref{fig:TN} and \ref{fig:TN3D}. Here, each individual tensor corresponds to the state of a small number of BC-bits, which are the original BC-bits for the external legs or new internal BC-bits. The internal edges of the tensor network correspond in path integral language to identifying the field configurations on the two BC bits joined by the edge and integrating over these. In quantum language, this corresponds to projecting the state of the pair of BC-bits onto a maximally entangled state. The full set of these projections contracts up the tensor network, defining what is known as a projected entangled pairs state (PEPS) (see, for example \cite{Orus:2013kga}).

There is clearly a great deal of freedom in how to build up a tensor network representation: we can choose where to place the new boundary components and we can choose the surfaces along which to break up $\tilde{H}$ to give the new internal BC-bits. We can also choose the geometry of $\tilde{H}$ from among the family of geometries related by conformal invariance. So, as expected, we can have many tensor networks that represent the same state. The networks here apparently have a closer connection to the geometry of the path-integral defining the state rather than the geometry of the space being encoded; such a connection was emphasized recently in \cite{Milsted:2018yur}. However, the recent work \cite{Caputa:2017urj,Caputa:2017yrh} suggests that by a particular optimization of the path-integral geometry over the geometries related by conformal invariance, the path-integral geometry actually becomes the geometry of the bulk spatial slice. Similarly, it may be that under some optimization of our tensor network description, we end up with networks similar to those in \cite{Pastawski:2015qua, Hayden:2016cfa} where the network geometry has a close connection to the bulk spatial geometry.

\subsubsection*{The limit of many BC-bits}

For simplicity, the various examples shown in the figures involve only a small number of BC-bits. In this case, the individual BC-bits might carry information about a significant portion of the dual geometry. However, our conclusions apply equally well when the number of BC-bits is taken to be very large, so long as the modifications leading from  $H$  to $\tilde{H}$ are kept small (e.g. each $\tilde{M}_i$ is still a large subset of $M_i$). In the limit where the BC-bits are all small compared to any scale associated with $M$ or the original CFT state, we expect that the individual bits carry almost no information about the geometry being represented by the collection of BC-bits, apart from the asymptotic behavior of the bulk fields at a single boundary location corresponding to the bit. Thus, the spacetime geometry is almost entirely encoded in the entanglement structure of multipart BC-bit system. In this system, by disentangling the bits, it is manifestly true that the corresponding spacetime disintegrates, as suggested for continuous CFTs in \cite{VanRaamsdonk:2010pw}. We emphasize that in the BC-bit system, the individual bits have no intrinsic location relative to one another,\footnote{Thinking of the BC-bits as pieces of a jigsaw puzzle, we are describing the same geometry whether the pieces are spread out on the table or the puzzle is completed.} so it is not only the radial direction of the spacetime that ``emerges.''

\subsubsection*{The speculative part}

Since the entanglement structure is playing such a key role here, and the individual bits carry almost no information about the interior of the spacetime, it is interesting to ask what properties of the BC-bit are really required here. If we replace the BC-bits with some other type of bit but keep the entanglement structure the same, does this still encode the same spacetime and gravitational physics? Initial inspection might suggest that this is too optimistic: the specific theory of gravity encoded in a CFT state has a certain number of dimensions and a certain set of fields in addition to the metric, and these are related to the dimensionality and the operator content of the CFT. If we try to replace BC-bits from one CFT with BC-bits from another CFT, possibly of a different dimensionality, it may seem that we can't possibly be describing the same gravitational physics since we are now dealing with a completely different theory of gravity.

But there is a way around this obstacle: the optimistic scenario could work if there is fundamentally only one theory of quantum gravity. This is in line with expectations from string theory \cite{Polchinski:1998rr}, where the study of string dualities suggest that various UV complete gravitational theories in various dimensions can be understood as descending through compactification and dualities from eleven-dimensional M-theory. Thus, even if the gravitational physics in one region of spacetime has the fields and interactions associated with a particular low-energy theory of gravity, the fields and interactions in another region of spacetime could correspond to a different low-energy theory if we have some transition region in between where the properties of the compactification manifold or the preferred duality frame change.

In our example where we replace BC-bits from one CFT with BC-bits associated with another CFT, we will clearly have different physics in the asymptotically AdS regions very close to the BC-bits (near the diamonds in figure \ref{fig:PathInt}f). But, moving inward in the radial directions, we may have a transition (e.g. with some compact dimensions changing shape or topology) such that inside a radial position associated with a boundary scale containing a large number of BC-bits, the physics is that of the spacetime described by the original CFT. If we replace the BC-bits with more general quantum systems (e.g. collections of qubits, or macaroni), it may be that the asymptotic region no longer has a geometrical description, but the same interior region emerges.

The idea that we could obtain a precise description of quantum gravitational physics starting from sufficiently many copies of an arbitrary quantum system is intriguing, but certainly does not follow from any of the arguments of this paper. Nevertheless, it is fascinating that the possible unity of gravitational theories as suggested by string theory leaves open the possibility of such an exact and universal entanglement-gravity duality.\footnote{This is in line with Susskind's slogan $GR=QM$ \cite{Susskind:2017ney}, though our discussion suggests that perhaps a better moniker would be $M=QM$ or, more simply, $Q=1$.}

\section*{Acknowledgements}

This work was supported in part by a Simons Investigator award and by a Simons Foundation collaboration grant. 

\bibliographystyle{JHEP}
\bibliography{BCbit}

\providecommand{\href}[2]{#2}\begingroup\raggedright\begin{thebibliography}{10}

\bibitem{Maldacena:2001kr}
J.~M. Maldacena, \emph{{Eternal black holes in anti-de Sitter}},
  \href{http://dx.doi.org/10.1088/1126-6708/2003/04/021}{\emph{JHEP} {\bf 04}
  (2003) 021}, [\href{https://arxiv.org/abs/hep-th/0106112}{{\tt
  hep-th/0106112}}].

\bibitem{Ryu:2006bv}
S.~Ryu and T.~Takayanagi, \emph{{Holographic derivation of entanglement entropy
  from AdS/CFT}},
  \href{http://dx.doi.org/10.1103/PhysRevLett.96.181602}{\emph{Phys. Rev.
  Lett.} {\bf 96} (2006) 181602},
  [\href{https://arxiv.org/abs/hep-th/0603001}{{\tt hep-th/0603001}}].

\bibitem{Swingle:2009bg}
B.~Swingle, \emph{{Entanglement Renormalization and Holography}},
  \href{http://dx.doi.org/10.1103/PhysRevD.86.065007}{\emph{Phys. Rev.} {\bf
  D86} (2012) 065007}, [\href{https://arxiv.org/abs/0905.1317}{{\tt
  0905.1317}}].

\bibitem{VanRaamsdonk:2009ar}
M.~Van~Raamsdonk, \emph{{Comments on quantum gravity and entanglement}},
  \href{https://arxiv.org/abs/0907.2939}{{\tt 0907.2939}}.

\bibitem{VanRaamsdonk:2010pw}
M.~Van~Raamsdonk, \emph{{Building up spacetime with quantum entanglement}},
  \href{http://dx.doi.org/10.1007/s10714-010-1034-0,
  10.1142/S0218271810018529}{\emph{Gen. Rel. Grav.} {\bf 42} (2010)
  2323--2329}, [\href{https://arxiv.org/abs/1005.3035}{{\tt 1005.3035}}].

\bibitem{Maldacena:2013xja}
J.~Maldacena and L.~Susskind, \emph{{Cool horizons for entangled black holes}},
  \href{http://dx.doi.org/10.1002/prop.201300020}{\emph{Fortsch. Phys.} {\bf
  61} (2013) 781--811}, [\href{https://arxiv.org/abs/1306.0533}{{\tt
  1306.0533}}].

\bibitem{VanRaamsdonk:2016exw}
M.~Van~Raamsdonk, \emph{{Lectures on Gravity and Entanglement}},  in
  \emph{{Proceedings, Theoretical Advanced Study Institute in Elementary
  Particle Physics: New Frontiers in Fields and Strings (TASI 2015): Boulder,
  CO, USA, June 1-26, 2015}}, pp.~297--351, 2017.
\newblock \href{https://arxiv.org/abs/1609.00026}{{\tt 1609.00026}}.
\newblock \href{http://dx.doi.org/10.1142/9789813149441_0005}{DOI}.

\bibitem{Pastawski:2015qua}
F.~Pastawski, B.~Yoshida, D.~Harlow and J.~Preskill, \emph{{Holographic quantum
  error-correcting codes: Toy models for the bulk/boundary correspondence}},
  \href{http://dx.doi.org/10.1007/JHEP06(2015)149}{\emph{JHEP} {\bf 06} (2015)
  149}, [\href{https://arxiv.org/abs/1503.06237}{{\tt 1503.06237}}].

\bibitem{Hayden:2016cfa}
P.~Hayden, S.~Nezami, X.-L. Qi, N.~Thomas, M.~Walter and Z.~Yang,
  \emph{{Holographic duality from random tensor networks}},
  \href{http://dx.doi.org/10.1007/JHEP11(2016)009}{\emph{JHEP} {\bf 11} (2016)
  009}, [\href{https://arxiv.org/abs/1601.01694}{{\tt 1601.01694}}].

\bibitem{Cardy:2004hm}
J.~L. Cardy, \emph{{Boundary conformal field theory}},
  \href{https://arxiv.org/abs/hep-th/0411189}{{\tt hep-th/0411189}}.

\bibitem{Skenderis:2008dh}
K.~Skenderis and B.~C. van Rees, \emph{{Real-time gauge/gravity duality}},
  \href{http://dx.doi.org/10.1103/PhysRevLett.101.081601}{\emph{Phys. Rev.
  Lett.} {\bf 101} (2008) 081601}, [\href{https://arxiv.org/abs/0805.0150}{{\tt
  0805.0150}}].

\bibitem{Botta-Cantcheff:2015sav}
M.~Botta-Cantcheff, P.~Martínez and G.~A. Silva, \emph{{On excited states in
  real-time AdS/CFT}},
  \href{http://dx.doi.org/10.1007/JHEP02(2016)171}{\emph{JHEP} {\bf 02} (2016)
  171}, [\href{https://arxiv.org/abs/1512.07850}{{\tt 1512.07850}}].

\bibitem{Marolf:2017kvq}
D.~Marolf, O.~Parrikar, C.~Rabideau, A.~Izadi~Rad and M.~Van~Raamsdonk,
  \emph{{From Euclidean Sources to Lorentzian Spacetimes in Holographic
  Conformal Field Theories}},
  \href{http://dx.doi.org/10.1007/JHEP06(2018)077}{\emph{JHEP} {\bf 06} (2018)
  077}, [\href{https://arxiv.org/abs/1709.10101}{{\tt 1709.10101}}].

\bibitem{Balasubramanian:2014hda}
V.~Balasubramanian, P.~Hayden, A.~Maloney, D.~Marolf and S.~F. Ross,
  \emph{{Multiboundary Wormholes and Holographic Entanglement}},
  \href{http://dx.doi.org/10.1088/0264-9381/31/18/185015}{\emph{Class. Quant.
  Grav.} {\bf 31} (2014) 185015}, [\href{https://arxiv.org/abs/1406.2663}{{\tt
  1406.2663}}].

\bibitem{Brown:2015bva}
A.~R. Brown, D.~A. Roberts, L.~Susskind, B.~Swingle and Y.~Zhao,
  \emph{{Holographic Complexity Equals Bulk Action?}},
  \href{http://dx.doi.org/10.1103/PhysRevLett.116.191301}{\emph{Phys. Rev.
  Lett.} {\bf 116} (2016) 191301},
  [\href{https://arxiv.org/abs/1509.07876}{{\tt 1509.07876}}].

\bibitem{Karch:2000gx}
A.~Karch and L.~Randall, \emph{{Open and closed string interpretation of SUSY
  CFT's on branes with boundaries}},
  \href{http://dx.doi.org/10.1088/1126-6708/2001/06/063}{\emph{JHEP} {\bf 06}
  (2001) 063}, [\href{https://arxiv.org/abs/hep-th/0105132}{{\tt
  hep-th/0105132}}].

\bibitem{Takayanagi:2011zk}
T.~Takayanagi, \emph{{Holographic Dual of BCFT}},
  \href{http://dx.doi.org/10.1103/PhysRevLett.107.101602}{\emph{Phys. Rev.
  Lett.} {\bf 107} (2011) 101602}, [\href{https://arxiv.org/abs/1105.5165}{{\tt
  1105.5165}}].

\bibitem{Chiodaroli:2011fn}
M.~Chiodaroli, E.~D'Hoker and M.~Gutperle, \emph{{Simple Holographic Duals to
  Boundary CFTs}}, \href{http://dx.doi.org/10.1007/JHEP02(2012)005}{\emph{JHEP}
  {\bf 02} (2012) 005}, [\href{https://arxiv.org/abs/1111.6912}{{\tt
  1111.6912}}].

\bibitem{Chiodaroli:2012vc}
M.~Chiodaroli, E.~D'Hoker and M.~Gutperle, \emph{{Holographic duals of Boundary
  CFTs}}, \href{http://dx.doi.org/10.1007/JHEP07(2012)177}{\emph{JHEP} {\bf 07}
  (2012) 177}, [\href{https://arxiv.org/abs/1205.5303}{{\tt 1205.5303}}].

\bibitem{Orus:2013kga}
R.~Orus, \emph{{A Practical Introduction to Tensor Networks: Matrix Product
  States and Projected Entangled Pair States}},
  \href{http://dx.doi.org/10.1016/j.aop.2014.06.013}{\emph{Annals Phys.} {\bf
  349} (2014) 117--158}, [\href{https://arxiv.org/abs/1306.2164}{{\tt
  1306.2164}}].

\bibitem{Milsted:2018yur}
A.~Milsted and G.~Vidal, \emph{{Tensor networks as path integral geometry}},
  \href{https://arxiv.org/abs/1807.02501}{{\tt 1807.02501}}.

\bibitem{Caputa:2017urj}
P.~Caputa, N.~Kundu, M.~Miyaji, T.~Takayanagi and K.~Watanabe, \emph{{Anti-de
  Sitter Space from Optimization of Path Integrals in Conformal Field
  Theories}},
  \href{http://dx.doi.org/10.1103/PhysRevLett.119.071602}{\emph{Phys. Rev.
  Lett.} {\bf 119} (2017) 071602},
  [\href{https://arxiv.org/abs/1703.00456}{{\tt 1703.00456}}].

\bibitem{Caputa:2017yrh}
P.~Caputa, N.~Kundu, M.~Miyaji, T.~Takayanagi and K.~Watanabe, \emph{{Liouville
  Action as Path-Integral Complexity: From Continuous Tensor Networks to
  AdS/CFT}}, \href{http://dx.doi.org/10.1007/JHEP11(2017)097}{\emph{JHEP} {\bf
  11} (2017) 097}, [\href{https://arxiv.org/abs/1706.07056}{{\tt 1706.07056}}].

\bibitem{Polchinski:1998rr}
J.~Polchinski, \emph{{String theory. Vol. 2: Superstring theory and beyond}}.
\newblock Cambridge Monographs on Mathematical Physics. Cambridge University
  Press, 2007,
  \href{http://dx.doi.org/10.1017/CBO9780511618123}{10.1017/CBO9780511618123}.

\bibitem{Susskind:2017ney}
L.~Susskind, \emph{{Dear Qubitzers, GR=QM}},
  \href{https://arxiv.org/abs/1708.03040}{{\tt 1708.03040}}.

\end{thebibliography}\endgroup

\end{document}